\documentclass[groupedaddress,
  aip,
  pop,
  amsmath,amssymb,
  reprint,
]{revtex4-1}
\usepackage{graphicx,color}
\usepackage{amsmath}
\usepackage{natbib}
\usepackage{epsfig}
\begin{document}

\title{\color{blue} Modified Frost formula for the mobilities of positive ions in their parent gases}

\author{Sergey A. Khrapak}
\email{Sergey.Khrapak@dlr.de}
\affiliation{Institut f\"ur Materialphysik im Weltraum, Deutsches Zentrum f\"ur Luft- und Raumfahrt (DLR), 82234 We{\ss}ling, Germany
}

\author{Alexey G. Khrapak}
 \email{khrapak@mail.ru}
 \affiliation{Joint Institute for High Temperatures, Russian Academy
of Sciences, 125412 Moscow, Russia}

\begin{abstract}
A modification to the semi-empirical Frost formula for the mobilities of singly charged positive ions in their parent gases is proposed. The modified expression demonstrates excellent agreement with experimental results for different ionized rare gases in a very extended range of reduced field strengths.   
\end{abstract}

\date{\today}

\maketitle

A very important characteristic of low-ionized plasmas is the mobility of ions under the action of an electric field. The concept of ion mobility is central to various phenomena, including ambipolar diffusion,~\cite{FrostPR1957,RaizerBook} sheath and boundary layer formation in gas discharges,~\cite{RiemannJPD1991,RiemannJPD1992,PhelpsJAP1994} macroparticle charging and the ion drag force in complex (dusty) plasmas,~\cite{BarnesPRL1992,KhrapakPRE2002,KhrapakPoP2005,FortovPR2005,ZobninPoP2008,KhrapakPRE2013,PuttscherPoP2014} as well as many others.

Often a simplified assumption is made that the ionic mobility is constant. This is not consistent with experimental measurements, which have shown that, for the most interesting case of atomic rare gas ions in their parent gases,  the mobility decreases with increasing electric field strength.~\cite{HornbeckPR1951,BiondiPR1954,Ellis1976,BasurtoPRE2000}  No general expression for the dependence of the ion mobility on the electric field is known. At the same time, several theoretical approximations have been suggested in the literature.~\cite{FrostPR1957,Wannier1953,Fahr1967,PattersonPRA1970,Hahn1972,LampePoP2012}               

Among the approximations that have been put forward, one of the most simple and convenient for practical use is the semi-empirical formula proposed by Frost:~\cite{FrostPR1957}     
\begin{equation}\label{Frost}
M=A\left[1+B\frac{E}{N}\right]^{-1/2}\frac{E}{N}.
\end{equation}
Here  $M$ denotes the ion drift velocity expressed in units of the ion thermal velocity, $M=u/v_T$, where $v_T=\sqrt{T/m}$, $T$ is the ion temperature, and $m$ is the ion mass ($M$ is often referred to as the thermal Mach number).  The ratio of electric field strength to the neutral gas number density $E/N$ is expressed in Townsend units (1 Td = $10^{-17}$ V cm$^2$).  The constants $A$ and $B$ are different for different gases. 

The physics behind Eq.~(\ref{Frost}) is as follows.~\cite{RaizerBook,Hahn1972} Elementary theories of ion drift in an electric field lead to
\begin{equation}\label{drift}
u=\frac{eE\tau}{m},
\end{equation}
where $e$ is the ion charge and $\tau$ is the characteristic average (momentum transfer) time between collisions with neutral atoms. This time can be expressed as
\begin{equation}
\tau=\left\langle\frac{1}{Nv\sigma(v)} \right\rangle,
\end{equation}  
where $\sigma(v)$ is the velocity-dependent momentum transfer cross section and $\langle...\rangle$ denotes an appropriate averaging over relative velocities $v$ between ions and neutrals. In the regime of weak electric fields and subthermal drifts, the averaging is essentially over thermal velocities and hence the collisional time $\tau\sim 1/Nv_T\sigma(v_T)$ is independent of drift velocity and electric field. As a results the drift velocity is directly proportional to $E/N$, irrespective of detailed shape of ion-neutral interactions. In the opposite limit of strong fields and highly superthermal drifts, the thermal contribution is negligible upon averaging and we get $\tau\sim 1/Nu\sigma(u)$. Here the dependence of the drift velocity on $E/N$ is determined by the nature of the ion-neutral interaction. For example, for a (perhaps not very realistic in the context of ion-neutral interactions)  inverse-power-law interaction potential, $\propto r^{-n}$, we get $\sigma(u)\propto u^{-4/n}$, provided $n\gg 1$. This leads to $u\propto (E/N)^{n/(2n-4)}$. In the limit of a rigid sphere interaction ($n\rightarrow\infty$) we obtain $u\propto (E/N)^{1/2}$. This latter limit is generally relevant, because at high energies the ion-neutral collision cross sections do approach constant asymptotes.~\cite{RaizerBook} For ions drifting in their parent gases the dominant collisions mechanism -- resonant charge-exchange -- has only weak logarithmic dependence on the relative velocity.~\cite{RaizerBook,Smirnov2015}         

Thus, the semi-empirical Frost formula (\ref{Frost}) represents just one particular simple way to interpolate between the limiting regimes of weak and strong electric fields. Generally, it is in a rather good agreement with experimental results on the drift velocities of rare gas ions in their parent gases. There is, however, some room for improvements. For example, the original Frost formula overestimates the Ar$^+$ ion mobility in Ar in the regime of weak electric field~\cite{RobertsonPRE2003,Cavity} (see e.g. Fig.~\ref{FigAr} below). 

The purpose of this Note is to demonstrate that a modest modification of the Frost formula allows us to reach excellent agreement with numerical data for different gases in the entire range $E/N$, where experimental data is available. The approximation we propose is    
\begin{equation}\label{mFrost}
M=A\left[1+\left(B\frac{E}{N}\right)^C\right]^{-1/2C}\frac{E}{N},
\end{equation}
where $C$ is the parameter of order unity. This formula is correct in the corresponding limits of weak and strong electric fields. For $C=1$ expressions (\ref{mFrost}) reduces to the conventional formula (\ref{Frost}).
 
\begin{figure}
\includegraphics[width=7.5cm]{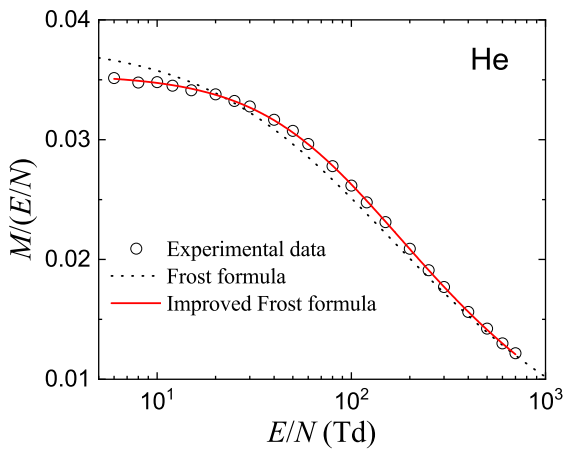}
\caption{Reduced mobility of He$^+$ ions in He gas as a function of the ratio of electric field strength to the neutral gas number density, $E/N$. The symbols are experimental data.~\cite{Ellis1976} The dotted curve corresponds to the original Frost formula~\cite{FrostPR1957} (\ref{Frost}), while the solid curve corresponds to its modification (\ref{mFrost}).  }
\label{FigHe}
\end{figure}

\begin{figure}
\includegraphics[width=7.5cm]{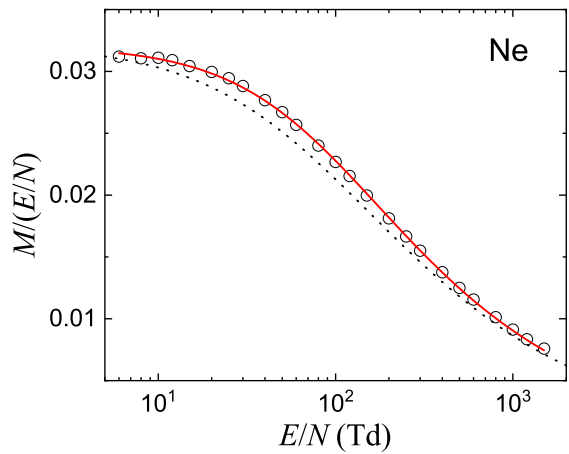}
\caption{Same as in Fig.~\ref{FigHe}, but for Ne$^+$ ions in Ne.}
\label{FigNe}
\end{figure}

\begin{figure}
\includegraphics[width=7.5cm]{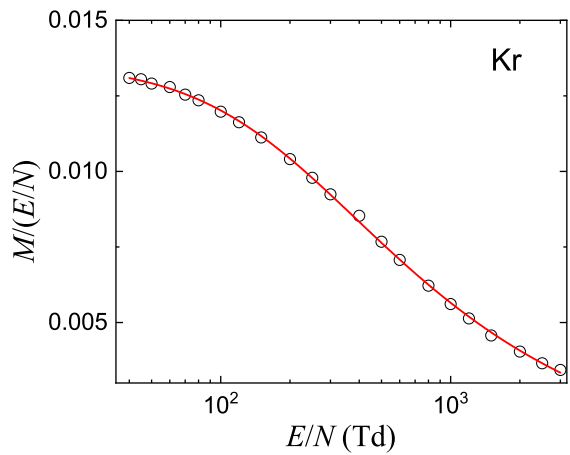}
\caption{Same as in Fig.~\ref{FigHe}, but for Kr$^+$ ions in Kr. Here only the modified formula is plotted, because $Kr$ gas was not considered in the original paper.~\cite{FrostPR1957}  }
\label{FigKr}
\end{figure}

Figures \ref{FigHe}-\ref{FigKr} demonstrate the comparison between the available experimental data on He$^+$, Ne$^+$, Ar$^+$, and Kr$^+$ drifts in their respective parent gases, the original Frost formula (\ref{Frost}), and the modified Frost formula (\ref{mFrost}). The proposed modification provides excellent agreement with the experimental results. The obtained fitting parameters $A$, $B$, and $C$ are summarized in Table~\ref{Tab1} [note that the parameters $A$ and $B$ are naturally somewhat different from those originally suggested to use in Eq.~(\ref{Frost})].  The last column lists the effective momentum transfer cross sections for slow subthermal ion drifts. These cross sections have been evaluated by combining the elementary formula for the ion drift (\ref{drift}) with the simple estimate $\tau^{-1}=Nv_T\sigma_{\rm eff}$ and using the corresponding parameter $A$ from the first column.  The modified Frost formula does not provide us with a simple analytical relation between the effective momentum transfer frequency and the reduced ion drift velocity.~\cite{KhrapakJPP2013} However, for practical numerical implementations its convenience is essentially the same as of the original formula, but the accuracy is higher.       

\begin{table}[!b]
\caption{\label{Tab1} Fitting parameters in the modified Frost formula (\ref{mFrost}).}
\begin{ruledtabular}
\begin{tabular}{lcccc}
System & $A$ (1/Td) & $B$ (1/Td) & $C$ & $\sigma_{\rm eff}$ ($10^{-14}$ cm$^2$) \\ \hline
He$^+$ in He &  0.0354 & 0.0118  & 1.355 & 1.09  \\
Ne$^+$ in Ne &  0.0321 & 0.0120 & 1.181  & 1.20 \\
Ar$^+$ in Ar &   0.0168 & 0.0070 & 1.238 & 2.30  \\
Kr$^+$ in Kr &   0.0136  & 0.0054 & 1.422 & 2.84 \\
\end{tabular}
\end{ruledtabular}
\end{table}

\begin{figure}
\includegraphics[width=7.5cm]{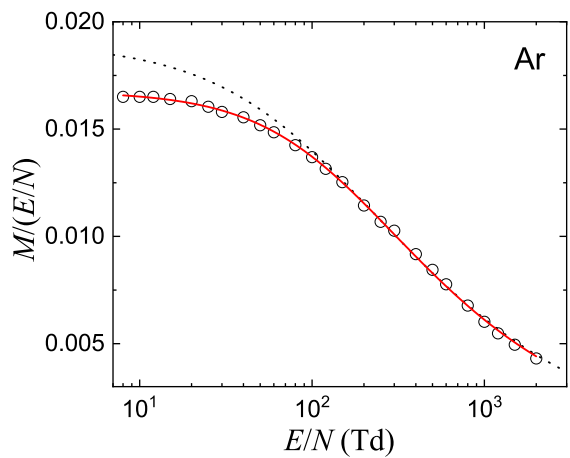}
\caption{Same as in Fig.~\ref{FigHe}, but for Ar$^+$ ions in Ar.}
\label{FigAr}
\end{figure}

As a final remark, the results presented here correspond to room temperature ions ($T=300$ K). Temperature dependence of the ion mobility has been investigated in Refs.~\onlinecite{GolyatinaLebedev2015,GolyatinaPPR2017}.

To conclude, we have suggested a modification to the original Frost formula for the mobilities of positive ions in their parent gases. The new expression is almost as simple as the original one, but  agrees considerably better with experimental results for different systems considered in the entire range of reduced electric field strength.

We would like to thank Victoriya Yaroshenko for careful reading of the manuscript.

\bibliographystyle{aipnum4-1}

\bibliography{Frost_References}

\end{document}